\documentclass{aastex63}

\usepackage{times} 
\usepackage{natbib}
\usepackage{amsmath}
\usepackage{amsfonts}
\usepackage{afterpage}
\makeatletter
\newcommand\footnoteref[1]{\protected@xdef\@thefnmark{\ref{#1}}\@footnotemark}
\usepackage[dvips]{epsfig}
\usepackage{amsmath}
\usepackage[flushleft]{threeparttable}
\usepackage{ragged2e}
\justifying
\submitjournal{ApJ}

\shorttitle{PAHs with aliphatic side groups}
\shortauthors{Buragohain et al.}
\graphicspath{{./}{figures/}}
\begin{document}
\title{DFT study on interstellar PAH molecules with aliphatic side groups}

\correspondingauthor{Mridusmita Buragohain}
\email{ms.mridusmita@gmail.com, mridusmita@astron.s.u-tokyo.ac.jp, amitpah@gmail.com}

\author{Mridusmita Buragohain}
\affiliation{Department of Astronomy, Graduate School of Science, The University of Tokyo, Tokyo 113-0033, Japan}

\author{Amit Pathak}
\affiliation{Department of Physics, Banaras Hindu University, Varanasi 221 005, India}

\author{Itsuki Sakon}
\affiliation{Department of Astronomy, Graduate School of Science, The University of Tokyo, Tokyo 113-0033, Japan}

\author{Takashi Onaka}
\affiliation{Department of Physics, Faculty of Science and Engineering, Meisei University}
\affiliation{Department of Astronomy, Graduate School of Science, The University of Tokyo, Tokyo 113-0033, Japan}




\begin{abstract}
Polycyclic Aromatic Hydrocarbon (PAH) molecules have been long adjudged to contribute to the frequently detected distinct emission features at 3.3, 6.2, 7.7, 8.6, 11.2 and 12.7~$\mu \rm m$ with weaker and blended features distributed in the 3$-$20~$\mu \rm m$ region. The comparatively weaker 3.4~$\mu \rm m$ emission feature has been attributed to have an aliphatic origin as carrier. PAH with aliphatic functional group attached to it is one of the proposed potential candidate carriers for the 3.4~$\mu \rm m$ emission band, however, the assignment of carrier is still enigmatic. In this work, we employ Density Functional Theory (DFT) calculation on a symmetric and compact PAH molecule; coronene (C$_{24}$H$_{12}$) with aliphatic side group to investigate any spectral similarities with observed features at 3$-$4~$\mu \rm m$. The side groups considered in this study are $-$H (hydrogenated), $-$CH$_3$ (methyl), $-$CH$_2-$CH$_3$ (ethyl) and $-$CH$=$CH$_2$ (vinyl) functional groups. Considering the possible presence of deuterium (D) in PAHs, we also include D in the aliphatic side group to study the spectral behavior. We present a detailed analysis of the IR spectra of these molecules and discuss possible astrophysical implications.
\end{abstract}


\section{Introduction}
Since the discovery of the Unidentified Infrared (UIR) emission bands, a quest for its carrier has led to a trail of ongoing research that has proposed several hypotheses to determine the structure of its carriers. \citet[]{Leger84} and \citet[]{Allamandola85, Allamandola89} independently proposed that the UIR bands arise due to the vibrational relaxation in the Polycyclic Aromatic Hydrocarbon (PAH) molecules on absorption of background photons \citep[for a recent review]{Tielens08}. Besides this, other popular hypotheses
are HAC (hydrogenated Amorphous Carbon) \citep[]{Jones90}, QCC (Quenched Carbonaceous Composites) \citep[]{Sakata87}, coal \citep[]{Papoular93} and MAON model (mixed aromatic/aliphatic 
organic nanoparticles) \citep[]{Kwok11, Kwok13a}.
Some interstellar PAHs giving rise to the UIR bands are suggested to carry aliphatic C$-$H bonds that give rise to the 3.4~$\mu \rm m$ emission feature near the aromatic C$-$H 3.3~$\mu \rm m$ feature. The 3.4~$\mu \rm m$ feature, though weak compared to the 3.3~$\mu \rm m$ one, is a well known feature and ubiquitously detected towards several astrophysical sources \citep[and references therein]{Li12}. Initially, the 3.3~$\mu \rm m$ feature was detected as a broad
emission feature \citep[]{Grasdalen76, Tokunaga90}. Later, higher resolution spectroscopy confirmed that the 3.3~$\mu \rm m$ emission feature indeed has sub-components that form a plateau centered near 3.45~$\mu \rm m$ and can extend up to 3.6~$\mu \rm m$ \citep[]{Geballe85}. These features are particularly observed
at 3.40, 3.46, 3.51, and 3.57~$\mu \rm m$ and are proposed to come from one of the followings (i) overtones and combinational bands of the fundamental C$-$C vibrational modes  \citep[]{Barker87, Geballe89, Muizon90}, (ii) C$-$H stretching modes in aliphatic hydrocarbons \citep[]{Duley81, Kondo12} or (iii) due to the presence of aliphatic functional group in a PAH molecule \citep[]{Geballe89, Muizon90, Joblin96, Sloan97}. The small value of I$_{3.4}$/I$_{3.3}$\footnote{the observed intensity ratio of 3.4~$\mu \rm m$ to 3.3~$\mu \rm m$ bands} \citep[]{Geballe89, Joblin96, Mori14} indicates that the UIR carriers are substantially aromatic.
\citet[]{Chiar2000, Chiar13} reported the observations of 3.4~$\mu \rm m$ band in absorption in the diffuse ISM along the line of sight towards the Galactic Center, which is supposed to come from submicron size dust grains. Using the 3.4~$\mu \rm m$ absorption band together with the 6.85 and 7.25~$\mu \rm m$ features, which also have an aliphatic origin, \citet[]{Chiar2000, Chiar13} discussed the aliphatic and aromatic absorption coefficients and proposed that Galactic hydrocarbon dust is highly aromatic.

Among several possibilities of attribution towards the 3.4~$\mu \rm m$ emission feature, we call this UIR feature aliphatic C$-$H stretching in a PAH molecule in this paper. The observed 3.3 and 3.4~$\mu \rm m$ features are useful to estimate the aromatic/aliphatic fraction in an interstellar PAH molecule giving rise to the UIR bands. \citet[]{Yang13, Yang16, Yang17a} proposed an upper limit of $\sim$2\% of aliphatic carbon atoms that might be attached to a PAH molecule and can account for the observed intensities of the 3.3 and 3.4~$\mu \rm m$ emission bands. They used the observed intensities of the 3.3 and 3.4~$\mu \rm m$ features and intrinsic band strengths of the theoretically computed 3.3 and 3.4~$\mu \rm m$ features (on a per unit C$-$H bond basis) to estimate the aliphatic to aromatic ratio in PAHs. 

The major objectives of this paper are to theoretically investigate
aliphatic C$-$H bonds in PAHs, which has not been studied very efficiently compared to the aromatic C$-$H characteristics in the past except for a few studies \citep[]{Bernstein96, Li12, Yang13, Yang16}, although features thought to arise from aliphatic C$-$H stretching are ubiquitously seen in the 3~$\mu \rm m$ spectrum. 
\citet[]{Geballe89, deMuizon86, Muizon90, Joblin96} proposed that 
additional satellite features near aromatic 3.3~$\mu \rm m$ might be a characteristic of aliphatic side group ($-$CH$_3$ or $-$CH$_2-$CH$_3$) attached to a PAH molecule. Another possible contributor is suggested as hydrogenated PAHs by \citet[]{Schutte93, Bernstein96}. By considering the attribution of aliphatic side group as well as hydrogenation, \citet[]{Pauzat99, Pauzat01}
calculated small PAHs with side groups and highly hydrogenated small PAHs, proposing that both attributions hold true to explain satellite features at 3.40 and 3.45~$\mu \rm m$. \citet[]{Sandford13} measured the laboratory infrared spectra of PAHs with excess peripheral H atoms and compared with astronomical spectra, suggesting that these molecules might be important in interstellar PAH population, though with a relatively low abundance. \citet[]{Maurya15} discussed the spectral characteristics of PAHs with unsaturated alkyl chains i.e., 
vinyl attached PAHs in relation to the UIR bands. The recent spectroscopic database published by \textit{NASA AMES} also
includes computationally calculated vibrational frequencies and intensities obtained from a few PAHs with 
side groups \citep[]{Bauschlicher18}.
In our study, we consider coronene (C$_{24}$H$_{12}$) as the parent molecule with aliphatic side chain of different forms, for example: hydrogenated, methyl, ethyl or unsaturated alkyl chains to understand the features at 3~$\mu \rm m$ region, both neutral and ionized.

On the other hand, \citet[]{Peeters04, Onaka14} and \citet[]{Doney16} identified that the aliphatic C$-$D feature at around 4.6~$\mu \rm m$ seem to be stronger than the aromatic C$-$D feature at around 4.4~$\mu \rm m$, suggesting that deuterium (D) may be more incorporated in aliphatic bonds than in aromatic, if they can be attributed to C$-$D stretching. In view of this, we include a D atom in the aliphatic side group to look for any possible signatures at 4.6$-$4.8~$\mu \rm m$ that may relate to the UIR bands. In this report, we present a detailed analysis of IR spectra of these molecules and discuss 
possible astrophysical implications. Since the observed UIR emission spectra are proposed to come from a mixture of all kinds of PAHs and the observed intensity of the 3.4~$\mu \rm m$ feature is comparatively weak, the fraction of PAHs with aliphatic side group may be quite small. The present study allows us to estimate
what fraction of the PAH mixture has aliphatic side chain and what fraction of D may be present in that aliphatic unit, if any.
\section{Computational Approach}
\begin{figure*}[ht]
\begin{center}
\vspace{-5em}
\includegraphics[width=14cm, height=10cm]{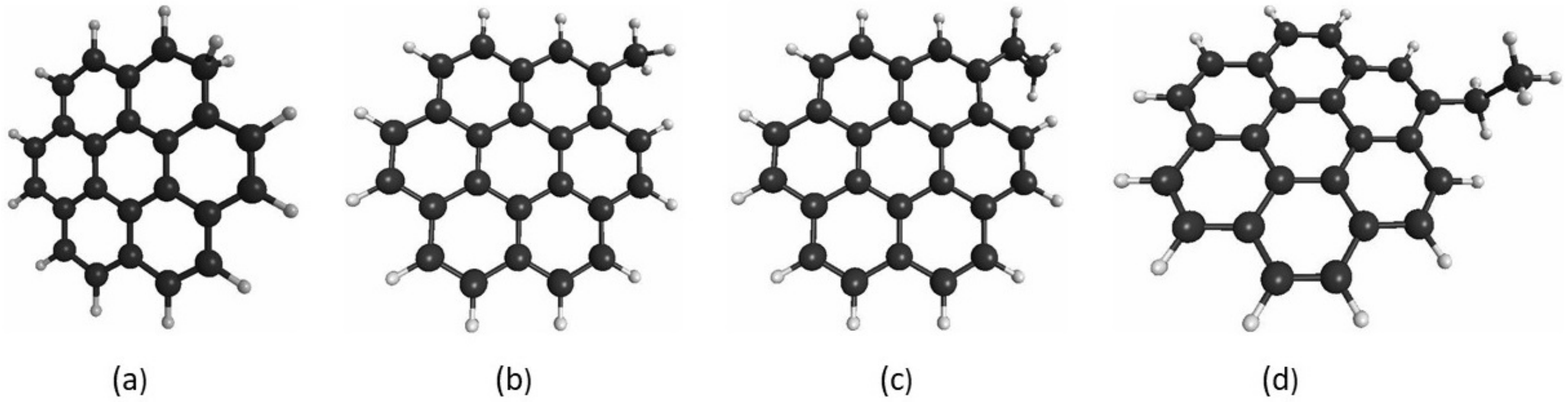}
\vspace{-10em}
\caption{Examples of PAH molecules with aliphatic side group studied in this work; (a) hydrogenated coronene, 
(b) coronene with a methyl side group, (c) coronene with a vinyl side group, (d) coronene with an ethyl side group}
\label{fig1}
\end{center}
\end{figure*}

Density Functional Theory (DFT) is by far the most successful approximation method to derive properties of a molecule, such as molecular structures, bond-lengths, vibrational frequencies, ionization energies, electric and magnetic properties, reaction paths, etc., based on determination of the electron density of the molecule \citep[]{DFT01}. This work reports DFT calculations on deuterated and deuteronated counterparts of PAHs with $-$H (hydrogenated), $-$CH$_3$ (methyl), $-$CH$_2-$CH$_3$ (ethyl) and $-$CH$=$CH$_2$ (vinyl) functional groups to determine the expected region of infrared features due to their characteristic vibrational modes. Coronene (C$_{24}$H$_{12}$), being a symmetric and compact molecule, has greater photostability against UV radiation. Therefore, we consider coronene molecule with functional groups attached to it for this study.
The structures of the sample molecules studied in this work are shown in Figure~\ref{fig1}. These type of molecules are expected to be present in benign environments of the ISM as they are easily prone to destruction in an intense UV irradiated region. 

For theoretical calculations, DFT in combination with a B3LYP/6-311G** has been
employed to optimize the molecular structure and calculate the frequency
of vibrational modes that are present in the corresponding molecule.
Also, mode dependent scaling factors have been used to scale the overestimated wavelengths.
This is described in detail in our previous reports 
\citep[]{Mridu15, Mridu18}. When the relative strengths of the modes obtained from 
DFT calculation are compared with those obtained from experiments,
the intensity of C$-$H stretching is found to be much larger
and an intensity scaling is also required apart from the existing wavelength scaling. 
Considering that the second order M\o{}ller-Plesset (MP2) perturbation theory with a large basis set (for example, 
MP2/6-311+G(3df, 3pd)) gives quite accurate oscillator strengths compared to B3LYP DFT, 
\citet[]{Yang17a} have derived a relation for MP2/6-311+G(3df, 3pd) and B3LYP/6-311+G** 
level of theories to scale the intensities of C$-$H stretching modes near $\sim$3~$\mu \rm m$ region. 
By using, A$_i$~{$\approx$}~0.6372 A$_j$, where A$_i$ and A$_j$ are the intensities of 
C$-$H stretching modes computed at the MP2/6-311+G(3df, 3pd) and
B3LYP/6-311+G** level, respectively, we can achieve good accuracy for band strength in the 3~$\mu \rm m$ region
by computing at an inexpensive level \citep[]{Yang17a}. The rest of the modes however, show better
matching in terms of band strength when compared with the experimentally obtained
spectra and do not require any scaling in their intensities.

For a valid and justifiable comparison with the observed UIRs, the theoretically computed absorption
spectrum needs to be transformed into an emission spectrum. An emission model is used
in order to obtain emission spectra of PAHs,
which may be {directly} compared with the observed emission features. 
Similar emission model has been used by \citet[]{Cook98, Pech02, Pathak08a}.
In the emission model, a PAH molecule is considered in a UV rich interstellar radiation field produced
by a source having an effective temperature of $\sim$~40,000~K. 
The PAH molecule absorbs this radiation of frequency $\nu$ with a cut off at 13.6 eV
and gets internally excited corresponding to an average temperature of about 1000 K. 
The absorption of the UV photon depends on the absorption
cross section ($\sigma_\nu$) of the particular PAH, which has been taken from
http://astrochemistry.ca.astro.it/database. The excited PAH then cools down by emitting in a cascade at 
frequencies corresponding to the vibrational modes of the PAH molecule. The emission model 
considers that within thermal approximation, the total energy of the excited PAH molecule is much greater than the average energy of an emitting mode \citep[]{Pech02, Pathak08a}. 
The emitted energy is integrated over the cooling range from 1000 K to 50 K with a decrease in internal energy by a temperature fall of 1 K. The emitted energy is calculated considering the rate of absorption of photons and added up over the whole distribution of photon absorption to produce the emission spectrum. The emitted energy and scaled wavelengths are plotted as Gaussian profiles with a FWHM of 30 cm$^{-1}$. The profile width depends on vibrational energy redistribution of the molecule and the chosen value of FWHM is typical for PAHs emitting in an interstellar environment \citep[]{Allamandola89}. Relative intensities (Int$\rm_{rel}$\footnote{Int$\rm_{rel}$=$\frac{\rm{absolute~intensity}}{\rm{maximum~absolute~intensity}}$}) are obtained by taking the ratio of all intensities to the maximum intensity.
\section{Results and Discussion}
\begin{figure*}[ht]
\begin{center}
\includegraphics[width=12cm, height=15cm]{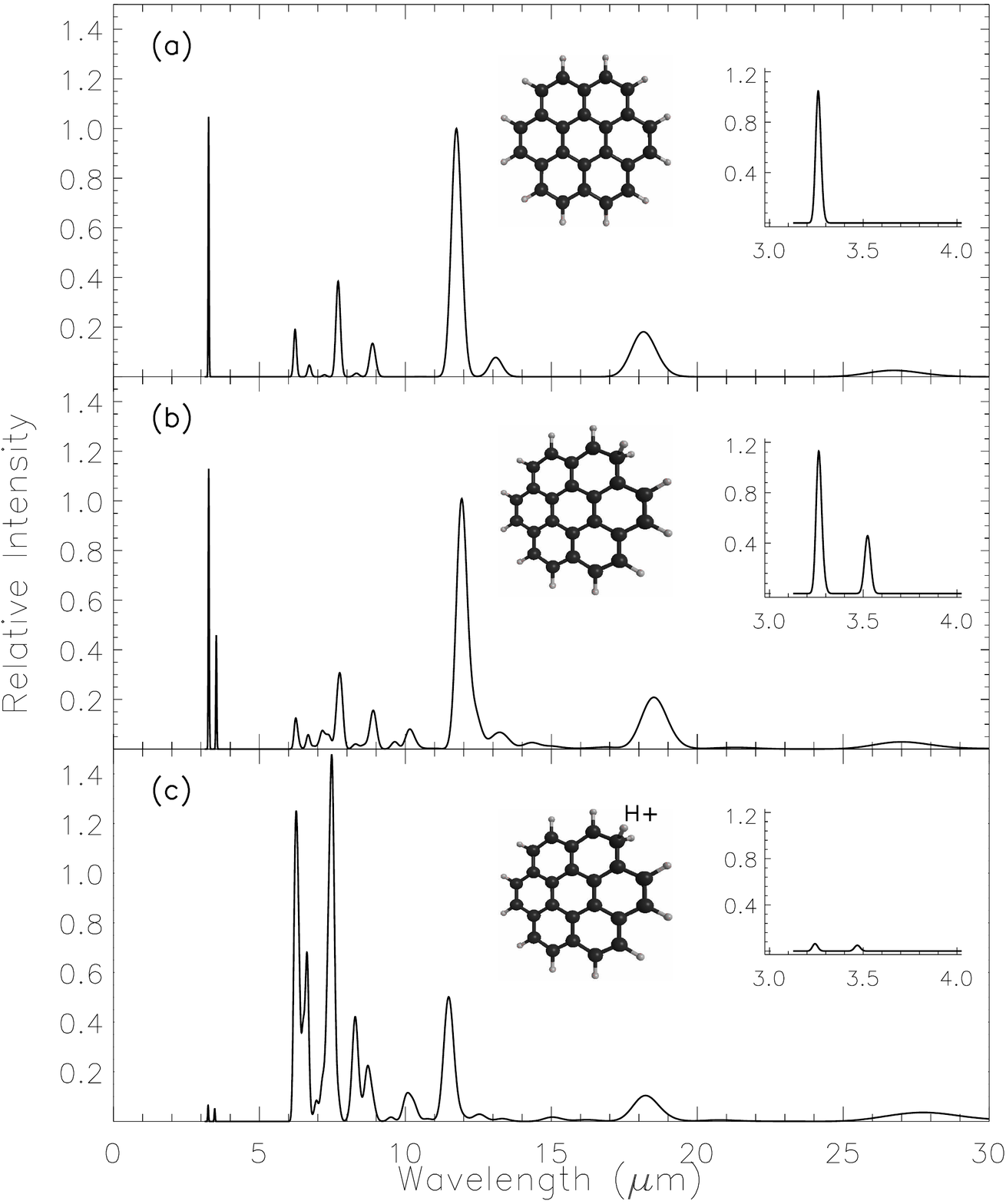}
\caption{Theoretical emission spectra of 3$-$20~$\mu \rm m$ region for (a) coronene (C$_{24}$H$_{12}$), 
(b) hydrogenated coronene (HC$_{24}$H$_{12}$), (c) Protonated coronene (HC$_{24}$H$_{12}^+$). An
enlargement of the 3$-$4~$\mu \rm m$ region is shown in insets that shows aliphatic and aromatic C$-$H stretching vibrational modes.}
\label{fig2}
\end{center}
\end{figure*}
Figure~\ref{fig2} shows a comparison of the 3$-$20~$\mu \rm m$ region of infrared
spectra of coronene (C$_{24}$H$_{12}$), hydrogenated coronene (HC$_{24}$H$_{12}$) and protonated coronene (HC$_{24}$H$_{12}^+$). The insets show the 3$-$4~$\mu \rm m$ region for the corresponding spectrum. A hydrogenated coronene, which is slightly aliphatic in nature shows almost similar features as that of a regular coronene molecule except distinctly around 3.5~$\mu \rm m$.
This is apparent as hydrogenated coronene is simply a coronene molecule with an additional hydrogen attached to it forming an aliphatic C$-$H bond at the addition site, the stretching of which introduces a feature 
at 3.5~$\mu \rm m$. The 3.5~$\mu \rm m$ feature in HC$_{24}$H$_{12}$ is a composite of two modes; symmetric C$-$H stretching at 3.52~$\mu \rm m$ and antisymmetric C$-$H stretching at 3.53~$\mu \rm m$ with 
relative intensities (Int$\rm_{rel}$) of 0.3 and 0.1, respectively. For protonated coronene, the intensity is {significantly} reduced near the 3~$\mu \rm m$ region. 
The 3.5~$\mu \rm m$ composite is slightly blueshifted and appears at $\sim$~3.47 with much weaker intensity (Int$\rm_{rel}\sim$~0.05). For both HC$_{24}$H$_{12}$ and HC$_{24}$H$_{12}^+$, there is no apparent
gap between symmetric and asymmetric C$-$H stretching modes. Another small difference from coronene molecule is that for hydrogenated and protonated coronene, a small plateau near 10~$\mu \rm m$ arises due to C$-$C$-$C$\rm_{in-plane}$ vibrational modes. 
This might be a reflection of the distorted symmetry of hydrogenated and protonated coronene 
from that of coronene after an additional H is being attached to coronene.
\begin{figure*}
\centering
\includegraphics[width=15cm, height=18cm]{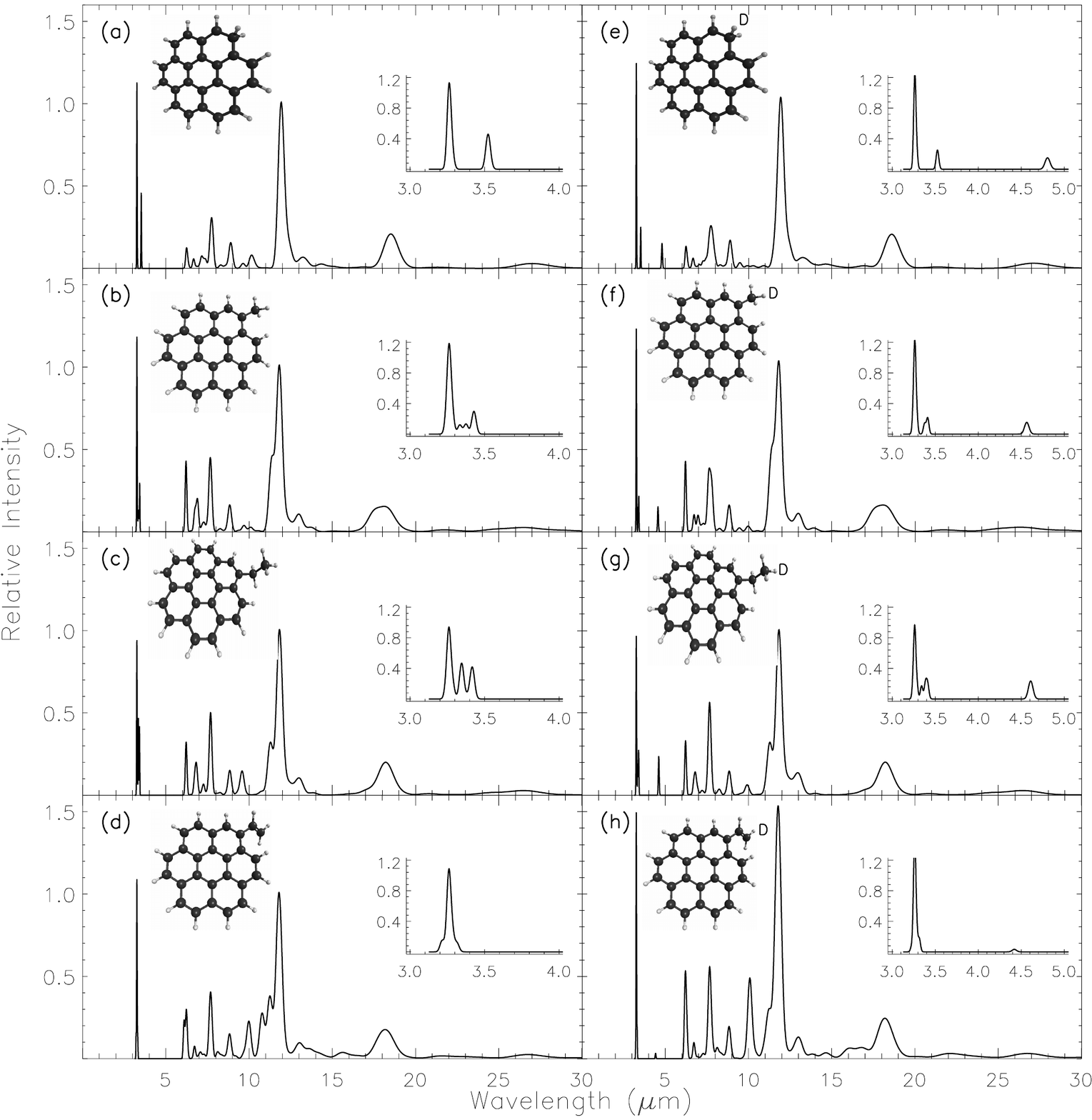}
\vspace{-4em}
\caption{Theoretical emission spectra of (a) HC$_{24}$H$_{12}$, 
(b) C$_{24}$H$_{11}-$CH$_3$, (c) C$_{24}$H$_{11}-$CH$_2-$CH$_3$, (d) C$_{24}$H$_{11}-$CH$=$CH$_2$,
(e) DC$_{24}$H$_{12}$, (f) C$_{24}$H$_{11}-$CH$_2$D, (g) C$_{24}$H$_{11}-$CH$_2-$CH$_2$D, (h) C$_{24}$H$_{11}-$CH$=$CHD. 
C$-$H stretching and C$-$D stretching vibrational modes are shown for each spectrum 
that arise between 3$-$4 and 3$-$5~$\mu \rm m$, respectively}
\label{fig3}
\end{figure*}

The theoretically obtained vibrational spectra of hydrogenated coronene (HC$_{24}$H$_{12}$), 
methyl$-$coronene (C$_{24}$H$_{11}-$CH$_3$), ethyl$-$coronene (C$_{24}$H$_{11}-$CH$_2-$CH$_3$)
and vinyl$-$coronene (C$_{24}$H$_{11}-$CH$=$CH$_2$) along with their deuterated counterparts\footnote{H or one of the H atoms {attached to the carbon atom} in the functional group is substituted with D} are presented in Figure~\ref{fig3}. With the size and form variation in the added aliphatic component, the symmetry of the 
molecule breaks down and new features appear in the spectrum. The lines in the spectrum are the manifestations of the characteristic vibrational 
modes occurring within a molecule. However, the effect of the presence of an aliphatic bond in the PAH molecule is only clearly evident from features near $\sim$3.5~$\mu \rm m$. Also, the presence of deuterium in PAH at the aliphatic site can be understood via C$-$D stretching near $\sim$~4.7~$\mu \rm m$. Hence in this report, we mainly focus on the discussion of 3$-$5~$\mu \rm m$ region to interpret any behaviour that corresponds to aliphatic {bonds} in a PAH molecule. There are features that represent other kinds of vibrational modes coming from the aliphatic bonds, but are too weak to be distinctly seen in the spectra. Apart from the aliphatic C$-$H stretching feature at $\sim$~3.4~$\mu \rm m$,
\citet[]{Li12} reported the appearance of a 6.85~$\mu \rm m$ feature that arises due to aliphatic C$-$H deformation. In our study, methyl$-$coronene (C$_{24}$H$_{11}-$CH$_3$) indeed shows a feature at 6.85~$\mu \rm m$ due to deformation of aliphatic C$-$H units attached to the PAH molecule. However, for other molecules, the position is slightly blueshifted and difficult to distinguish from combinational modes of C$-$C stretching and C$-$H in plane vibrations. Also,
the feature is comparatively less intense for hydrogenated coronene and vinyl$-$coronene. 
\subsection{Aliphatic nature of neutral PAHs in the \texorpdfstring{3$-$5~$\mu \rm m$~}~region}
For hydrogenated coronene (HC$_{24}$H$_{12}$), the aliphatic C$-$H stretching feature distinctly appears near 3.5~$\mu \rm m$ while for methyl$-$coronene (C$_{24}$H$_{11}-$CH$_3$) and ethyl$-$coronene (C$_{24}$H$_{11}-$CH$_2-$CH$_3$), the same feature spreads and is slightly blueshifted towards 3.4~$\mu \rm m$ extending down to 
3.3~$\mu \rm m$. For hydrogenated coronene, there is no apparent gap between symmetric and asymmetric C$-$H stretching modes, while for other side groups, features present at shorter wavelengths than the gap are usually attributed to asymmetric C$-$H stretching and those at longer wavelengths are attributed to symmetric C$-$H stretching modes. \citet[]{Chiar2000, Chiar13} similarly identify shorter and longer features to those asymmetric and symmetric stretching modes from methyl (CH$_3$) and methylene group (CH$_2$) in the absorption spectra towards diffuse ISM. While the aliphatic~features
are blueshifted with inclusion of side groups 
as compared to hydrogenation, these still maintain to separate out from the comparatively stronger aromatic features at $\sim$~3.26~$\mu \rm m$. The asymmetric  and symmetric C$-$H stretching groups appear at average positions of 3.36~$\mu \rm m$ and 3.42~$\mu \rm m$, respectively, for both methyl ($-$CH$_3$) and ethyl ($-$CH$_2-$CH$_3$) side groups and are close to the values given by \citet[]{Chiar2000, Chiar13} for methyl group
(3.376~$\mu \rm m$ and 3.474~$\mu \rm m$, respectively).

However, for vinyl substituted coronene (C$_{24}$H$_{11}-$CH$=$CH$_2$), aliphatic C$-$H bonds in the $-$CH$=$CH$_2$ side group do not show aliphatic stretching near 3.4~$\mu \rm m$ and instead fall in the aromatic region, i.e., near 3.3~$\mu \rm m$. In fact, a small 3.21~$\mu \rm m$ (Int$\rm_{rel}\sim$~0.1) arises due to the asymmetric stretching of aliphatic C$-$H bond and even one of the aromatic C$-$H stretching modes appear at 3.22~$\mu \rm m$ (Int$\rm_{rel}\sim$~0.07). Similar results were presented by \citet[]{Yang17a} and \citet[]{Maurya15} for vinyl substituted PAHs. \citet[]{Pauzat99} also find a 3.2~$\mu \rm m$ signature in methylenic ($=$CH$_2$) sub group and propose that since there is no report for detection of the {UIR} features at around 3.2~$\mu \rm m$, the abundance of the vinylic sub group should be quite small. \citet[]{Maurya15} also show additional features near 6.2~$\mu \rm m$ due to vinyl C$=$C stretch  and near 10~$\mu \rm m$ due to vinyl CH$_2$ wag and twist motions in vinyl substituted PAHs. In this study, we find peaks near 6.1~$\mu \rm m$ and 10.8~$\mu \rm m$ in case of vinyl substituted coronene. 

With increasing size of the functional groups {attached to PAH}, the number of aliphatic C$-$H bonds
increases and so are the vibrational modes due to aliphatic C$-$H stretching {covering the $\sim$~3.31$-$3.53~$\mu \rm m$ region.} For example: there are three features at 3.33 (Int$\rm_{rel}\sim$~0.12), 3.37 (Int$\rm_{rel}\sim$~0.13)
and 3.43~$\mu \rm m$ (Int$\rm_{rel}\sim$~0.3) that arise due to the stretching of aliphatic C$-$H bonds in 
methyl coronene (C$_{24}$H$_{11}-$CH$_3$). This causes an increase in the integrated intensity {of the aliphatic C$-$H stretching bands}. However, inclusion of additional hydrogen or side group does not affect the integrated intensity of aromatic C$-$H stretching 
modes at $\sim$3.26~$\mu \rm m$ and that of aromatic C$-$H stretching modes is always greater than that of aliphatic for single hydrogenation or single side group substitution. The ratio of the integrated intensity of the 3.3~$\mu \rm m$ band to that of 3.4$-$3.5~$\mu \rm m$ bands, {i.e., $\frac{\rm Int_{3.3}}{\rm Int_{3.4-3.5}}$\footnote{{To make it simpler, we call the aromatic C$-$H stretching in this paper as the 3.3~$\mu \rm m$ and aliphatic C$-$H stretching feature as the 3.4$-$3.5~$\mu \rm m$ though the aliphatic C$-$H stretching vibrational mode starts from 3.31~$\mu \rm m$ for some side groups.}}} decreases with increasing number of aliphatic C$-$H bonds ($\sim$2.9, $\sim$2.7, $\sim$1.3 for $-$H, $-$CH$_3$, $-$CH$_2-$CH$_3$, respectively.)
The positions and relative intensities (Int$\rm_{rel}>$~0.05 only) of aliphatic/aromatic C$-$H stretching for the sample molecules are presented in Table~\ref{tab1}.

\citet[]{Pauzat99, Pauzat01} proposed that hydrogenation contributes to the longer wavelength side of 3.46$-$3.50~$\mu \rm m$, while features in the shorter wavelength side of 3.38$-$3.45~$\mu \rm m$ can be attributed to methyl substitution. Though our study considers single hydrogenation and single side group substitution, the results are quite similar to \citet[]{Pauzat99, Pauzat01} who calculated strongly hydrogenated PAHs. This study finds that features at $\sim$~3.52~$\mu \rm m$ and $\sim$~3.33$-$3.43$\mu \rm m$ can be attributed to single hydrogenation and single methyl group ($-$CH$_3$) substitution, respectively. In addition, we find that ethyl group ($-$CH$_2-$CH$_3$) can equally be important to cause features at $\sim$~3.34$-$3.42~$\mu \rm m$.

The bond dissociation energy (BDE) to remove the attached aliphatic side group is also given in Table~\ref{tab1}. In the harsh condition of the astronomical source, the chemical pathways might be different and form an intermediate product before losing 
\begin{longrotatetable}
\begin{table}[htb]
\movetableright=-2cm
 \caption{Intensities and positions of the C$-$H/C$-$D stretching mode in PAHs (neutral) with aliphatic components\label{tab1}}
\scalebox{0.75}{
 \begin{tabular}[c]{c|c|c|c|c||c|c|c|c|c}
 \hline \hline
 Molecules & \footnotesize{Mode} & Peak wavelength & Normalized intensity & BDE & Molecules & \footnotesize{Mode} & Peak wavelength  & Normalized intensity & BDE \\ 
& \footnotesize{assignment} & ($\mu \rm m$) & Int$\rm_{rel}$ & (eV) & & \footnotesize{assignment} & ($\mu \rm m$) & Int$\rm_{rel}$ & (eV) \\ \hline
&C$-$H stretching \footnotesize{(arom)} & 3.26 & 0.15 & & & C$-$H stretching \footnotesize{(arom)} & 3.26 & 0.17 & \\ 
 & do & 3.26 & 0.56 & &  & do & 3.26 & 0.61 & \\
 & do & 3.26 & 0.25 & & & do & 3.26 & 0.28 & \\
 \footnotesize{HC$_{24}$H$_{12}$}  & do & 3.26 & 0.07 & 1.21 & \footnotesize{DC$_{24}$H$_{12}$}  & do & 3.26 & 0.08 & 1.21  \\
& do & 3.28 & 0.07 & & & do & 3.28 & 0.08 & \\
& do & 3.29 & 0.06 & & & do & 3.29 & 0.07 & \\
 &C$-$H stretching  \footnotesize{(aliph)}  & 3.52 & 0.35 & & & C$-$H stretching \footnotesize{(aliph)}  & 3.52 & 0.25 & \\
& do  & 3.53 & 0.11 & & & C$-$D stretching \footnotesize{(aliph)}  & 4.81 & 0.15 & \\ \hline
&C$-$H stretching \footnotesize{(arom)} & 3.24 & 0.14 & &  & C$-$H stretching \footnotesize{(arom)} & 3.24 & 0.15 & \\
 & do & 3.26 & 0.28  & &  & do & 3.26 & 0.29 & \\
  & do & 3.26 & 0.61 & &  & do & 3.26 & 0.65 & \\
& do & 3.26 & 0.11 & & & do & 3.26 & 0.12 & \\
 \footnotesize{C$_{24}$H$_{11}-$CH$_3$} & do & 3.27 & 0.06 & 4.12 & \footnotesize{C$_{24}$H$_{11}$-CH$_2$D} & do & 3.27 & 0.06 & 4.05 \\
& do & 3.28 & 0.06 & & & do & 3.28 & 0.07 & \\
& do & 3.28 & 0.1 & & & do & 3.28 & 0.09 & \\
& C$-$H stretching \footnotesize{(aliph)} & 3.33 & 0.12 & & & C$-$H stretching \footnotesize{(aliph)} & 3.38 & 0.14 & \\
 & do  & 3.38 & 0.13 & & &  do & 3.41 & 0.21 &\\
& do  & 3.43 & 0.30 & & & C$-$D stretching \footnotesize{(aliph)}  & 4.56 & 0.15 & \\ \hline
&C$-$H stretching \footnotesize{(arom)} & 3.24 & 0.12 & & & C$-$H stretching \footnotesize{(arom)} & 3.24 & 0.13 & \\
 & do & 3.26 & 0.24 & & & do & 3.26 & 0.26 & \\
  & do & 3.26 & 0.47 & &  & do & 3.26 & 0.48 &  \\
& do & 3.26 & 0.07 & & & do & 3.26 & 0.06 & \\
 \footnotesize{C$_{24}$H$_{11}-$CH$_2-$CH$_3$} & do & 3.26 & 0.05 & 3.94 & \footnotesize{C$_{24}$H$_{11}-$CH$_2-$CH$_2$D} & do & 3.26 & 0.05 & 3.86 \\
& do & 3.29 & 0.11 & & & do & 3.27 & 0.05 & \\
& C$-$H stretching \footnotesize{(aliph)} & 3.34 & 0.18 & & & do & 3.29 & 0.11 & \\
& do & 3.35 & 0.31  & & & C$-$H stretching \footnotesize{(aliph)} & 3.34 & 0.17 & \\
 & do   & 3.41  & 0.18 & &  &  do & 3.39 & 0.18 & \\
& do  & 3.42 & 0.25 & & & do & 3.41 & 0.18 & \\ 
&  &  &  & & & C$-$D stretching \footnotesize{(aliph)}  & 4.61 & 0.23 & \\ \hline
&C$-$H stretching \footnotesize{(aliph)} & 3.21 & 0.1 & & & C$-$H stretching \footnotesize{(arom)} & 3.22 & 0.06 & \\
 & C$-$H stretching \footnotesize{(arom)} & 3.22 & 0.07 & & & C$-$H stretching \footnotesize{(aliph)} & 3.25 & 0.07 & \\
  & do & 3.26 & 0.25 & & & C$-$H stretching \footnotesize{(arom)} & 3.26 & 0.32 & \\
& do & 3.26 & 0.58 & & & do & 3.26 & 0.78 & \\
 \footnotesize{C$_{24}$H$_{11}-$CH$=$CH$_2$}& do & 3.26 & 0.13 & 4.51 & \footnotesize{C$_{24}$H$_{11}-$CH$=$CHD} & do & 3.26 & 0.18 & 4.44 \\
& do & 3.27 & 0.08 & & & do & 3.27 & 0.11 & \\
&  do & 3.28 & 0.09 & & & do & 3.28 & 0.05 & \\
& C$-$H stretching \footnotesize{(aliph)} & 3.29 & 0.05 & & & do & 3.28 & 0.12 & \\
& do & 3.31 &  0.1 & & & C$-$H stretching \footnotesize{(aliph)} & 3.31 & 0.17 & \\
\hline
\end{tabular}}
 \begin{tablenotes}
 \small
 \item Note: `arom' stands for aromatic
 \item \ \ \ \ \ \ \ \ \ \ `aliph' stands for aliphatic 
 \item \ \ \ \ \ \ \ \ \ \ `BDE' stands for bond dissociation energy for hydrogen at an aliphatic site or in a side group
 \item \ \ \ \ \ \ \ \ \ \ \ Int$\rm_{rel}$ smaller than 0.05 are not listed
\end{tablenotes}
\end{table}
\end{longrotatetable}
\noindent
the attached side group completely. It is found that among all the sample molecules considered in this study, the extra hydrogen in hydrogenated coronene is most loosely bound to the parent molecule compared to other side groups and it requires only 1.21 eV to remove the extra hydrogen at the aliphatic site from the parent molecule. This is understood as HC$_{24}$H$_{12}$ possess an open-shell structure and considered less stable compared to others, which have closed-shell structures. Vinyl side group ($-$CH$=$CH$_2$) is more strongly bound compared to others and 4.51 eV is needed to remove vinyl group from the parent molecule.

The deuterated counterparts of the molecules show almost similar features as those of aliphatic PAHs without D, 
but with an additional feature near 4.6~$\mu \rm m$ due to the stretching of the aliphatic C$-$D bond. The 3.5~$\mu \rm m$ is also seen due to the stretching of the aliphatic C$-$H bond, which is present at the same site as aliphatic C$-$D bond. Deuteration or deuterium included aliphatic side group does not shift the position of aromatic and aliphatic C$-$H stretching significantly. Also, it does not affect much the integrated intensity of aromatic C$-$H stretching at 3.3~$\mu \rm m$. It decreases the integrated intensity of aliphatic C$-$H stretching at 3.4~$\mu \rm m$ as compared to molecules without D substitution. However, with increasing size of deuterated side group (for example, from $-$D to $-$CH$_2-$CH$_2$D), the integrated intensity of aliphatic C$-$H stretching is still increasing except for $-$CH$=$CHD side group. For deuterated coronene (DC$_{24}$H$_{12}$), C$-$D stretching is seen
at 4.8~$\mu \rm m$ (Int$\rm_{rel}\sim$~0.15), and it is shifted to 4.6~$\mu \rm m$ (Int$\rm_{rel}\sim$~0.15)
with larger side group ($-$CH$_2$D). For $-$CH$_2-$CH$_2$D group, the feature remains at the same position with a slight increase in the intensity (Int$\rm_{rel}\sim$~0.23). The exception is C$_{24}$H$_{11}-$CH$=$CHD, for which C$-$D stretching does not appear at 4.6~$\mu \rm m$ and falls at the aromatic region of C$-$D stretching, i.e., at 4.4~$\mu \rm m$ (Int$\rm_{rel}\sim$~0.03). This is analogous to vinyl substituted coronene (C$_{24}$H$_{11}-$CH$=$CH$_2$), in which characteristic frequencies of aliphatic C$-$H stretching corresponding to the vinyl side group ($-$CH$=$CH$_2$) move to the aromatic region. In C$_{24}$H$_{11}-$CH$=$CHD, D substitution also enhances features in the $\sim$~6$-$12~$\mu \rm m$ region as compared to features in C$_{24}$H$_{11}-$CH$=$CH$_2$. The BDE of the side groups with D and without D are almost the same, which indicates that both forms of side groups can be equally important in terms of stability.

Though the aliphatic nature of the PAH molecule can be distinctly understood from features at $\sim$~3.4$-$3.5 and $\sim$~4.6~$\mu \rm m$ due to aliphatic C$-$H stretching and aliphatic C$-$D stretching, respectively, there are some indirect effects due to the inclusion
of an aliphatic side group to coronene. For example: usually, coronene has duo C$-$H groups\footnote{A duo C$-$H group, also referred as doubly-adjacent C$-$H group, is a group with one neighbouring adjacent C$-$H group on the same ring} at its periphery, the out-of-plane vibrations of which give intense features at $\sim$12~$\mu \rm m$. However, with one of the peripheral H atoms being replaced with $-$CH$_3$/$-$CH$_2$D, $-$CH$_2-$CH$_3$/$-$CH$_2-$CH$_2$D or $-$CH$=$CH$_2$/$-$CH$=$CHD side groups, the C$-$H group near the replacement site of these side 
groups becomes solo C$-$H group\footnote{A solo C$-$H group, also referred as non-adjacent C$-$H group, is
a group with no neighbouring adjacent C$-$H group on the same ring}, as a result of which a s
olo C$-$H out-of-plane vibrational mode (from the solo C$-$H group) appears at $\sim$~11.3~$\mu \rm m$. However, this situation is particular for molecules that originally has duo C$-$H groups and addition of an aliphatic side group 
like $-$CH$_3$ converts one of the duo C$-$H groups into a solo C$-$H group. The transformation of duo C$-$H group to solo C$-$H group is also a possibility for a partially dehydrogenated C$_{24}$H$_{12}$
\citep[]{Mridu18}. In such a case, the C$-$H out-of-plane vibrational mode of a converted solo C$-$H group 
is found to appear at $\sim$~11.5~$\mu \rm m$.
\begin{figure*}
\centering
\includegraphics[width=15cm, height=18cm]{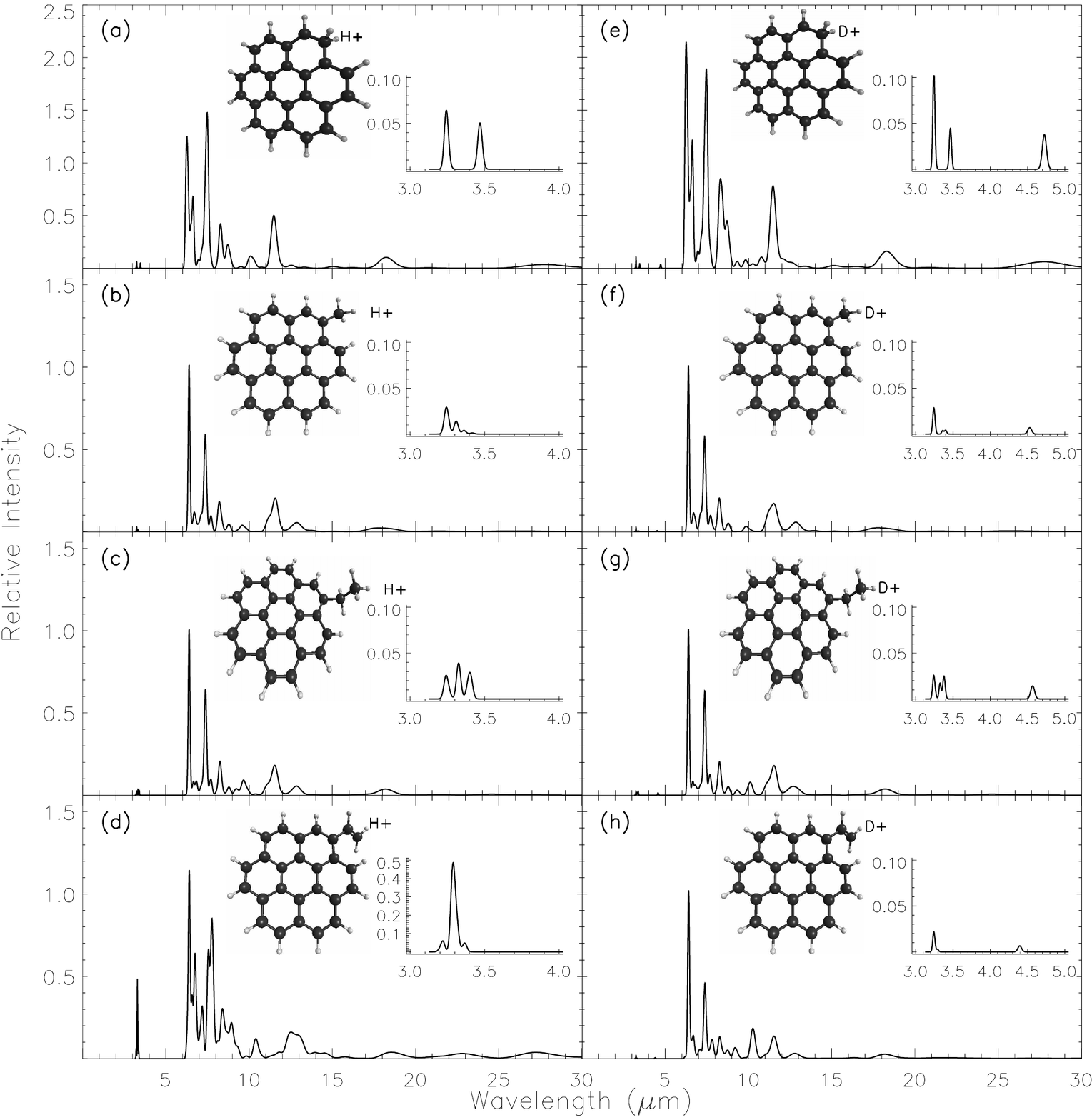}
\vspace{-4em}
\caption{Theoretical emission spectra of (a) HC$_{24}$H$_{12}^+$, 
(b) C$_{24}$H$_{11}-$CH$_3^+$, (c) C$_{24}$H$_{11}-$CH$_2-$CH$_3^+$, (d) C$_{24}$H$_{11}-$CH$=$CH$_2^+$,
(e) DC$_{24}$H$_{12}^+$, (f) C$_{24}$H$_{11}-$CH$_2$D$^+$, (g) C$_{24}$H$_{11}-$CH$_2-$CH$_2$D$^+$, 
(h) C$_{24}$H$_{11}-$CH$=$CHD$^+$. C$-$H stretching and C$-$D stretching vibrational modes
are shown for each spectrum that arise between 3$-$4 and 3$-$5~$\mu \rm m$, respectively}
\label{fig4}
\end{figure*}
\subsection{Aliphatic nature of protonated/deuteronated PAHs in the \texorpdfstring{3$-$5~$\mu \rm m$} region}
In testing conditions of the ISM, molecules are likely to be ionized on being illuminated by the UV radiation. For such a case, the ionized forms of sample PAH molecules 
is also considered in our study. In figure~\ref{fig4},
we present the theoretically obtained spectra of the same molecules (as shown in figure~\ref{fig3}), but in their ionized forms that correspond to protonated (HPAH$^+$) or deuteronated PAHs (DPAH$^+$). The structures of protonated and deuteronated PAHs are shown in figure~\ref{fig4} (as insets) where proton (H$^+$) or deuteron (D$^+$) has replaced H or one of H atoms in the aliphatic side group.
Protonation or deuteronation introduces new features in the existing spectra. These new features might arise due to the merging of several fundamental modes producing combinational modes. Like any other ionized PAHs, protonated and deuteronated PAHs show dominant features in the 6$-$10~$\mu \rm m$ region and weak features in other regions including the 3$-$5~$\mu \rm m$ region. The C$_{24}$H$_{11}-$CH$=$CH$_2^+$, however, shows 
a comparatively significant aromatic 3.3~$\mu \rm m$ feature (Figure~\ref{fig4}d) unlike other ionized PAHs.
Ionization causes a small wavelength blueshifting of aliphatic C$-$H/C$-$D stretching modes in 
protonated/deuteronated PAHs as compared to their respective neutral forms, while the aromatic C$-$H stretching almost retains its original position. For example: aliphatic C$-$H stretching for HC$_{24}$H$_{12}$ appears at $\sim$~3.52~$\mu \rm m$, whereas the same vibrational mode in HC$_{24}$H$_{12}^+$ appears at $\sim$~3.46~$\mu \rm m$. Similarly, aliphatic C$-$D stretching mode shifts from 4.81~$\mu \rm m$ 
to 4.73~$\mu \rm m$ in DC$_{24}$H$_{12}^+$ upon ionization of DC$_{24}$H$_{12}$. This shifting is not evident in $-$CH$_3^+$/$-$CH$_2$D$^+$, $-$CH$_2-$CH$_3^+$/$-$CH$_2-$CH$_2$D$^+$ 
or $-$CH$=$CH$_2^+$/$-$CH$=$CHD$^+$ side group substitution. 

The absolute and relative intensities of aromatic C$-$H stretching as well as aliphatic C$-$H/C$-$D stretching modes are considerably low in ionized molecules. \citet[]{Pauzat01} reported drastic intensity change only for aromatic C$-$H stretching and not for aliphatic C$-$H stretching for light hydrogenation for the corresponding ions. The intensity change for aliphatic C$-$H stretching is only apparent in case of strong hydrogenation. However, even for single hydrogenation and single side group substitution, our study finds a considerable intensity change, even for aliphatic C$-$H stretching modes as a result of ionization, although the change in intensity for aliphatic C$-$H stretching is small compared to aromatic C$-$H stretching. For example: for protonated coronene (HC$_{24}$H$_{12}^+$), the integrated absolute intensity due to aromatic and aliphatic C$-$H stretching is 15.76 km/mol and 11.308 km/mol, respectively, while for neutral form i.e., HC$_{24}$H$_{12}$, the values are 136.14 km/mol (aromatic C$-$H stretching) and 47.266 km/mol (aliphatic C$-$H stretching), respectively. The intensity of aliphatic C$-$D stretching also decreases upon ionization.
In case of DC$_{24}$H$_{12}^+$, the absolute intensity for aliphatic C$-$D stretching drops to 4.52 km/mol (Int$\rm_{rel}\sim$~0.04) from its respective value of 12.12 km/mol (Int$\rm_{rel}\sim$~0.15) in DC$_{24}$H$_{12}$. Table~\ref{tab1} only lists the distinct modes (Int$\rm_{rel}>$~0.05) in the 3$-$5~$\mu \rm m$ region and hence does not include any modes from protonated/deuteronated PAHs.

In our previous reports, we considered deuteronated PAHs (DPAH$^+$) and discussed important characteristics in relation to observed astronomical features \citep[]{Mridu15}. Here, we consider not only simple deuteronated PAHs, but other forms in which deuteron forms the part of the functional group to become a member of aliphatic PAHs. The aliphatic C$-$D stretching feature in DC$_{24}$H$_{12}^+$ (4.73~$\mu \rm m$) is blueshifted with larger functional groups
of which D$^+$ forms the part. The feature is seen at 4.5~$\mu \rm m$ (Int$\rm_{rel}\sim$~0.01) and 4.6~$\mu \rm m$ (Int$\rm_{rel}\sim$~0.01) for C$_{24}$H$_{11}-$CH$_2$D$^+$ and 
C$_{24}$H$_{11}-$CH$_2-$CH$_2$D$^+$, respectively. For coronene with deuteron included vinyl group (C$_{24}$H$_{11}-$CH$=$CHD$^+$), the stretching mode of the aliphatic C$-$D bond appears at the aromatic region of 4.4~$\mu \rm m$  (Int$\rm_{rel}\sim$~0.01).
\vspace{-12em}
\begin{figure*}[ht]
\centering
\includegraphics[width=15cm, height=18cm]{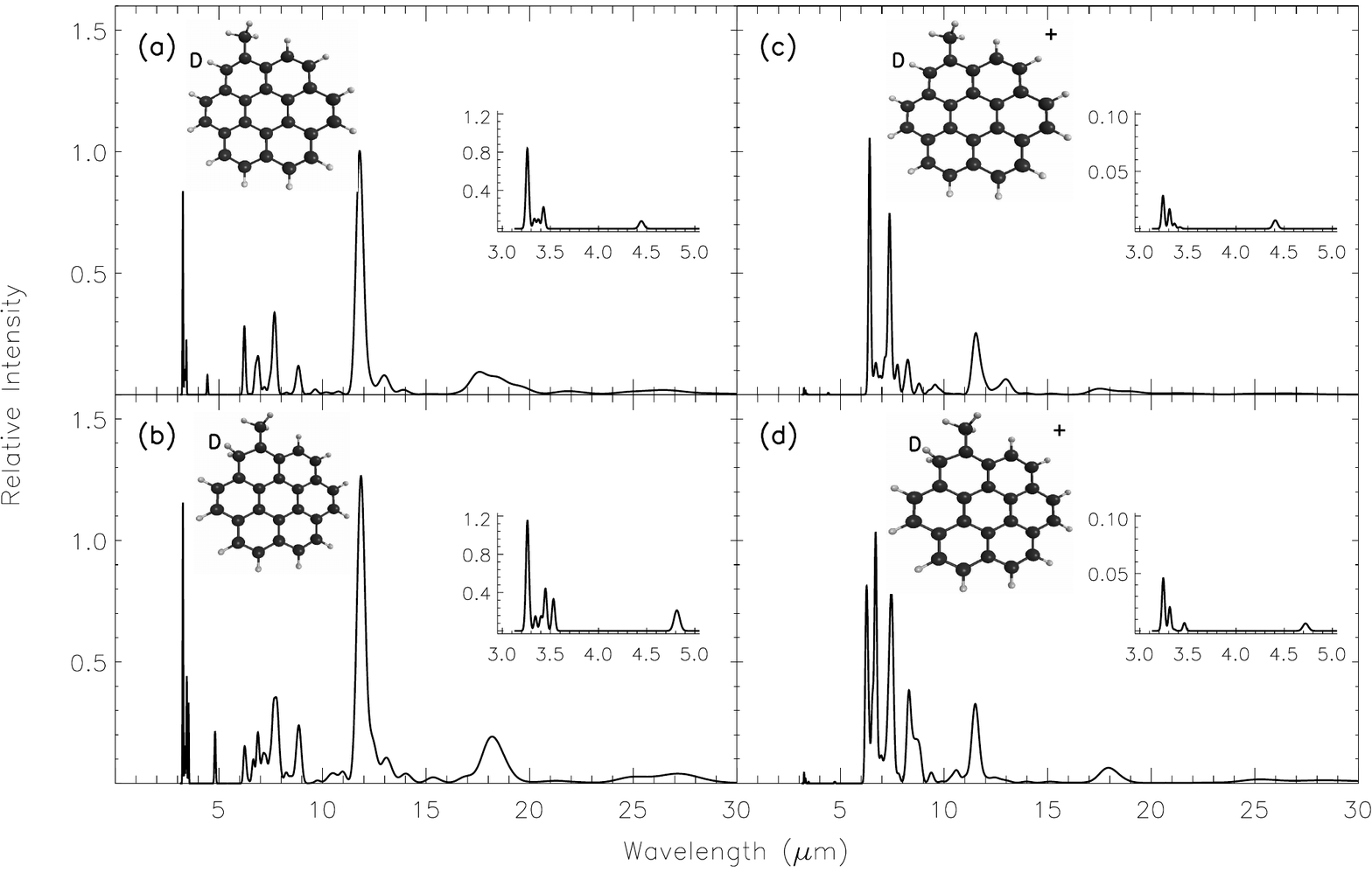}
\vspace{-15em}
\caption{Theoretical emission spectra of (a) C$_{24}$H$_{10}$D$-$CH$_3$, (b) C$_{24}$H$_{11}$D$-$CH$_3$, 
(c) C$_{24}$H$_{10}$D$-$CH$_3^+$, (d) C$_{24}$H$_{11}$D$-$CH$_3^+$. 3$-$5~$\mu \rm m$ region 
is shown for each spectrum.}
\label{fig5}
\end{figure*}

\citet[]{Pauzat99, Pauzat01} reported that for cations, the aliphatic C$-$H stretching spreads from 3.35~$\mu \rm m$ to 3.49~$\mu \rm m$. This study presents similar attribution upon ionization of the sample molecules. 
For our sample ionized molecules, features are seen 
at $\sim$~3.31$-$3.47~$\mu \rm m$ due to aliphatic C$-$H stretching. Considering the slight blueshifting upon ionization, for example: for hydrogenated coronene, the aliphatic C$-$H stretching shifts from 3.52~$\mu \rm m$ in neutral to 3.47~$\mu \rm m$ in protonated coronene, ionized molecules can be important in terms of positions for a few UIR bands in the aliphatic regime. However, the intensity for all cations is too small to accommodate the UIR bands at this position.

The interstellar D may not essentially be part of the aliphatic side groups in interstellar PAHs. Instead, it can be present 
independently anywhere on the PAH structure, although observations may suggest that D may be in aliphatic sites with possible detection of a weak 4.65~$\mu \rm m$, but no 4.4~$\mu \rm m$ feature \citep[]{Peeters04, Onaka14, Doney16}. In figure~\ref{fig5}, we have shown the infrared spectra of two such possibilities where deuterium is present in both aromatic and aliphatic site of 
the PAH molecule and does not form part of the methyl ($-$CH$_3$) functional group. Their ionized counterparts (deuteronated PAHs) are also considered. C$_{24}$H$_{10}$D$-$CH$_3$ carries the C$-$D bond at the aromatic site and shows the aromatic C$-$D stretching
at 4.4~$\mu \rm m$, aliphatic C$-$H stretching (in the $-$CH$_3$ group) at 3.4~$\mu \rm m$ and aromatic C$-$H stretching at 3.3~$\mu \rm m$, with significant intensities. In C$_{24}$H$_{11}$D$-$CH$_3$, the C$-$D bond is aliphatic in nature, the stretching of 
which produces the 4.8~$\mu \rm m$ feature in the spectra.
An aliphatic C$-$H bond is present in the addition site of D, which is not part of the $-$CH$_3$ functional group. The aliphatic C$-$H stretching at the addition site of D is slightly redshifted to 3.5~$\mu \rm m$ compared to aliphatic C$-$H stretching in the $-$CH$_3$ group at 3.4~$\mu \rm m$. Ionization of these molecules (C$_{24}$H$_{10}$D$-$CH$_3^+$, C$_{24}$H$_{11}$D$-$CH$_3^+$) considerably reduces the intensities of all features at 3.3, 3.4, 4.4 and 4.8~$\mu \rm m$ as shown in figure~\ref{fig5}~(c-d).
\section{Astrophysical Implications}
As mentioned in the introduction, several observations show the presence of a comparatively weak 3.4~$\mu \rm m$ emission feature in the neighborhood of the stronger 3.3~$\mu \rm m$ feature \citep[]{Geballe85, Geballe89, Muizon90, Joblin96, Mori14} and is attributed to possess an aliphatic origin as the carrier. Observations also show a weak plateau centered at $\sim$~3.45~$\mu \rm m$ consisting of minor features with small peaks
at 3.4, 3.46, 3.51 and 3.57~$\mu \rm m$ near
dominant 3.29~$\mu \rm m$ \citep[]{Geballe85, Geballe89, Roche96, Sloan97}. \citet[]{Geballe89} observed spatial
variation of these weak features relative to the 3.29~$\mu \rm m$ in the Orion bar and Red Rectangle and attributed the 3.4 and 3.51~$\mu \rm m$ features either to i) hot bands of the C$-$H stretching or to ii) C$-$H stretching in the aliphatic side group attached to PAH.
The 3.46 and 3.57~$\mu \rm m$ features were proposed to arise due to overtones and combination bands of low frequency C$-$C stretching modes instead. Using high resolution IR absorption spectra obtained from small sized PAH molecules, 
\citet[]{Maltseva18} propose that the 3~$\mu \rm m$ plateau 
is dominated by anharmonic effects.
\citet[]{Joblin96} studied the 3.29 and 3.4~$\mu \rm m$ features in spectra of two reflection nebulae; NGC 1333 SVS3 and NGC 2023. The spatially resolved 3~$\mu \rm m$ spectroscopy at different positions of the reflection nebulae showed that at certain positions, the 3.4~$\mu \rm m$ band becomes intense and the variation in the intensity ratio of 3.29 and 3.4~$\mu \rm m$ bands can be interpreted in term of the photochemical evolution of alkylated PAHs, particularly methylated PAHs \citep[]{Joblin96}. A decrease in the peak intensity ratio of the 
3.4~$\mu \rm m$ to 3.3~$\mu \rm m$ bands (I$_{3.4}$/I$_{3.3}$) was reported with an increase in the strength of the UV radiation field \citep[]{Joblin96}. 
Similarly, with increasing distance from the stellar source, an increase in the I$_{3.4}$/I$_{3.3}$ was seen in
\citet[]{Geballe89}. These observations suggest that with decreasing distance towards the illuminating source,
aliphatic bonds in the UIR band carriers undergo decomposition to become substantially aromatic that can account for the observed ratio of I$_{3.4}$/I$_{3.3}$ \citep[]{Joblin96, Pilleri15}. The aromatization of carbonaceous dust may occur as a result of UV photo processing \citep[]{Jones17}. \citet[]{Mori14} also showed that I$_{3.4-3.6~\mu \rm m}$/I$_{3.3~\mu \rm m}$ decreases with the increase of the PAH ionization degree, which
indicates that aliphatic C$-$H bonds are less resilient compared to aromatic C$-$H in the ionized gas-dominated region. The strength of the 3.4~$\mu \rm m$ is usually seen weak compared to the 3.3~$\mu \rm m$ in celestial sources. This may be because aliphatic bonds, if present in a PAH molecule, can easily be destroyed compared to its aromatic bonds by strong UV radiation in the ISM.

In this report, we consider PAHs with aliphatic side 
group ($-$H, $-$CH$_3$, $-$CH$_2-$CH$_3$, $-$CH$=$CH$_2$) to investigate aliphatic signatures in PAHs. 
The evolution or destruction of PAH molecule in any astronomical sources depends on the 
intensity of the UV flux, H density, H loss rate, rehydrogenation rate, and so on in that particular source and these
parameters vary from source to source. Depending on the size of the molecule, 
their fate is determined by chemical and physical processes. Considering
these parameters, PAHs of $\sim$~20$-$50 C atoms have been proposed as the UIR band carriers,
which undergo photochemical evolution as a function of the distance from the exciting star and 
can account for some of the observed features near 3.4~$\mu \rm m$ \citep[]{Geballe89}. 
The size of the surviving PAH and degree of hydrogenation or side group
substitution susceptible to photochemical erosion/evolution 
can vary from source to source and even within the same source at different distances.

\begin{figure*}[t]
\centering
\begin{center}
\includegraphics[width=12cm, height=12cm]{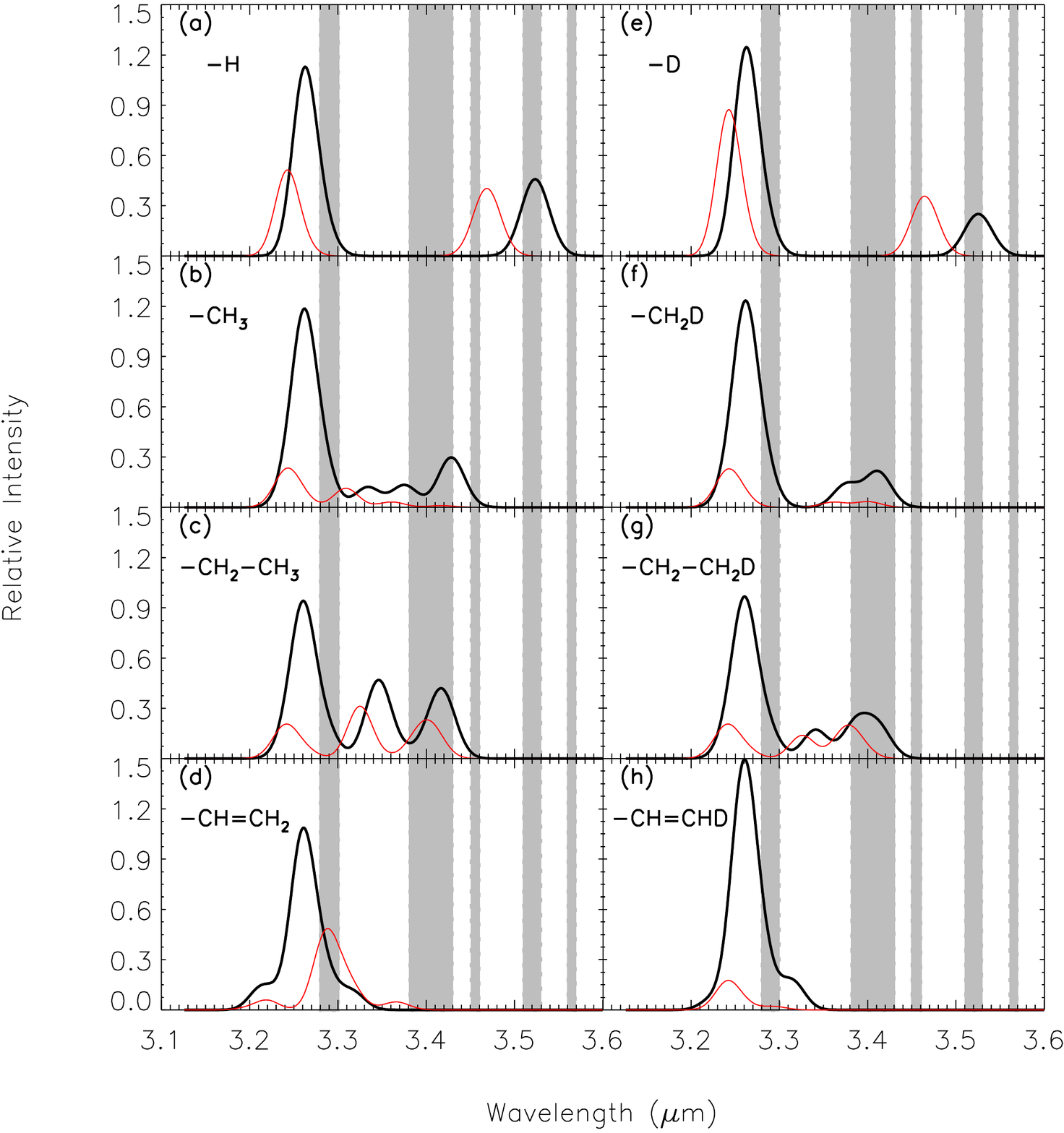}
\vspace{-5em}
\caption{The black line shows the emission spectra of neutral molecules with side groups, while the red line is for the corresponding ions. The side groups are labeled in the figure. 
The gray areas show the positions for observed emission features taken from several references \citep[]{Geballe85, Geballe89,Muizon90}. For all side groups except for vinyl group
($-$CH$=$CH$_2$), the red line for ions is scaled by a factor of 8 to make it visible.}
\label{fig6}
\end{center}
\end{figure*}

Due to the stretching of the aromatic and aliphatic C$-$H bonds in our sample molecules with side groups, a variety of features in the 3$-$4~$\mu \rm m$ region appear, which are close to some of the observed features in terms of band positions. Figure~\ref{fig6} shows the calculated positions of aromatic and aliphatic C$-$H stretching modes together with the positions for observed features in the 3$-$4~$\mu \rm m$ region marked by the gray area, which are taken from several references \citep[]{Geballe85, Geballe89, Muizon90}. 
Calculated positions for both neutral and ionized molecules are also
shown. As the ionized molecules produce faint features in this region, the features for ions are scaled by a factor of 8. For all side groups for neutral molecules, the peak position of the strongest feature due to aromatic C$-$H stretching appears at $\sim$~3.26~$\mu \rm m$, which is slightly at the shorter side compared to
the observed peak at $\sim$~3.29~$\mu \rm m$. This might be due to uncertainties in theoretical calculation, which has a typical uncertainty of $\sim$~10 wavenumbers ($\sim$~0.01~$\mu \rm m$ at 3.3~$\mu \rm m$). The aliphatic C$-$H stretching are close to a few observed features in the $\sim$~3.4$-$3.5~$\mu \rm m$, however, some features are not coinciding very well. For the hydrogenated/deuterated group ($-$H/$-$D), the aliphatic C$-$H stretching feature at $\sim$~3.52~$\mu \rm m$ is coinciding with the observed position and this may be attributed to the longer wavelength side of aliphatic domain in terms of band position. The methyl group ($-$CH$_3$/$-$CH$_2$D) shows consistency with observed features at $\sim$~3.42~$\mu \rm m$ and may be attributed to features at the shorter wavelength side. The ethyl group ($-$CH$_2-$CH$_3$/$-$CH$_2-$CH$_2$D) shows a match at the observed position of $\sim$~3.40$-$3.42~$\mu \rm m$, however, another distinct feature appears at $\sim$~3.35~$\mu \rm m$, which has never been observed firmly. This might be a general trend for PAHs with ethyl group as we find similar feature in case of ethyl pyrene and ethyl ovalene, smaller and larger than ethyl coronene in size, respectively from our DFT calculation. \citet[]{Pauzat99} computed similar features centered at 3.36~$\mu \rm m$ and 3.425~$\mu \rm m$ with significant intensities for ethyl pyrene. The methyl group also shows aliphatic C$-$H stretching vibrational modes at similar positions,
but with comparatively lower intensities. Observations only suggest a plateau at 3.35 micron that extends to 3.6~$\mu \rm m$ and no sharp feature of equivalent intensity as $\sim$~3.40$-$3.42~$\mu \rm m$ has been identified at $\sim$~3.35~$\mu \rm m$. This makes ethyl group comparatively a less likely UIR candidate in comparison to methyl group. It should be noted, however, that this study is restricted to coronene molecule and this implication may not be explicitly true for all size and form of PAHs. For the vinyl group 
($-$CH$=$CH$_2$/$-$CH$=$CHD), the aliphatic C-H stretching features are seen in the aromatic region of $\sim$~3.26~$\mu \rm m$ and hence may be excluded as carriers for the 3.4$-$3.5~$\mu \rm m$ UIR bands. Though ionized molecules show match for a few observed features, say $\sim$~3.40~$\mu \rm m$ for ethyl group and $\sim$~3.45~$\mu \rm m$ for hydrogenated PAHs, however, their extremely weak intensities make them unrealistic as probable carriers for observed bands at these wavelengths.

\citet[]{Yang13, Yang17a} calculated the band strength ratio of 
aliphatic C$-$H stretching to aromatic C$-$H stretching to be $\sim$~1.69 for neutral and $\sim$~3.48 for cations. 
The values are obtained from mono methyl derivatives of small PAHs. 
Even with larger side groups (ethyl, propyl and butyl), which are attached to neutral molecules, \citet[]{Yang16} computed a similar ratio of $\sim$~1.97. They further compared this ratio obtained from neutral molecules with the observed intensity ratio of 3.4~$\mu \rm m$ to 3.3~$\mu \rm m$ ($\sim$0.12) and estimated the ratio of number of C atoms in aliphatic unit to that of aromatic ring to be 
$\sim$~2\%, suggesting that the UIR emitters are predominantly aromatic. We obtained similar band strength ratios, A$_{3.4}$/A$_{3.3}$\footnote{A$_{3.4}$/A$_{3.3}$=$\frac{\rm band~
strength~of~the~aliphatic~C-H~stretching~(per~aliphatic~C-H~bond)}{\rm band~
strength~of~the~aromatic~C-H~stretching~(per~aromatic~C-H~bond)}$} from hydrogenated and methyl/ethyl substituted PAHs. The obtained average values from our calculation are 1.67 for neutral and 3.65 for cations. As the measured values of A$_{3.4}$/A$_{3.3}$ from our sample molecules are similar to those obtained by \citet[]{Yang13, Yang16, Yang17a}, a similar fraction of C atoms in the form of aliphatic unit in the UIR carriers may be expected. Even if we consider deuterium included aliphatic side group, which decreases the absolute intensity of aliphatic C$-$H stretching with no change in the absolute intensity from aromatic C$-$H stretching, the A$_{3.4}$/A$_{3.3}$ does not vary much and the proposal of the UIR carriers being highly aromatic is still valid.
\section{Conclusion}
The vibrational spectra of PAH molecule (coronene) with aliphatic side groups are studied in this work. The substitution of side groups in PAH does not alter the intensity and position of aromatic C$-$H stretching and the attribution of the 3.3~$\mu \rm m$ feature to aromatic C$-$H bonds still holds. The inclusion of an aliphatic side group to PAH causes aliphatic C$-$H stretching vibrational modes and the corresponding intensities and positions are dependent on the form of the side group. A distinct 3.5~$\mu \rm m$ feature appears due to aliphatic C$-$H stretching in hydrogenated coronene, which is then shifted towards 3.4~$\mu \rm m$ when larger side groups are considered. With the increasing number of aliphatic C$-$H bonds in larger side groups, more features at nearby positions arise, which might at least account for some of the observed bands in 3.4$-$3.5~$\mu \rm m$ region. The distinct emission features in the 3.4$-$3.5~$\mu \rm m$ region are mainly observed at 3.40, 3.46, 3.51, and 3.57~$\mu \rm m$. Though the calculated positions
of aliphatic C$-$H stretching for side groups are close to the observed positions, a few features are not coinciding very well. Among the sample molecules studied in this work, hydrogenated molecules might at least be responsible for observed features near 3.51~$\mu \rm m$, while PAHs with methyl ($-$CH$_3$) group might be accountable for features near 3.4~$\mu \rm m$. Despite a good match for a feature at $\sim$~3.42~$\mu \rm m$, the appearance of an additional distinct band at $\sim$~3.35~$\mu \rm m$ in ethyl group ($-$CH$_2-$CH$_3$)
makes it a weak UIR candidate carrier in this region.  Vinyl ($-$CH$=$CH$_2$) substituted PAH molecules also may not be a very likely candidate for any of the observed features near 3.4$-$3.6~$\mu \rm m$ region as aliphatic features of such molecule are seen at the aromatic region. For the remaining two observed features 
at 3.46 and 3.57~$\mu \rm m$, our sample molecules do not show any convincing match and the origins for such features may not be associated with aliphatic side group and rather be coming due to overtones and combination bands of C$-$C stretching vibrations.   

The interstellar deuterium (D) may replace H atom at aliphatic sites and form deuterated/deuteronated PAHs. Such molecules
show a weak 4.6~$\mu \rm m$ due to aliphatic C$-$D stretching and might be potential candidate carriers for the recently observed band at 4.6~$\mu \rm m$. Deuteration do not affect the previous attribution much, which makes deuterium included side groups equally important as other side groups without D. Ionization reduces the intensity of aliphatic/aromatic C$-$H stretching by a considerable amount, which excludes ionized molecules from being potential
candidate carriers for this region. Because of computational limitations, this study is restricted to coronene (C~$\sim$~24) and a detailed study including a variety of PAHs with isomers is needed to draw a concrete conclusion about the aliphatic nature of the
UIR carriers. Based upon the calculated wavelength position and intensity, laboratory experiments of interesting molecules would be useful for further study of the identification of the UIR band carriers. Inclusion of methyl and ethyl side group shows a few weak vibrational modes at nearby positions of $\sim$3.4~$\mu \rm m$, which have not been firmly detected and instead a broad plateau with underlying feature has been suggested by observations. High spectral resolution of JWST will offer us prospective to look into details of this plateau which would provide constraints on the nature of the UIR carriers.
%
\section{Acknowledgements}
MB thanks the Japan Society for Promotion of Science (JSPS) for awarding her research fellowship and grant.
AP acknowledges financial support from DST EMR grant, 2017 (SERB-EMR/2016/005266) 
and thanks the Inter-University Centre for Astronomy and Astrophysics, Pune for associateship.
The authors also acknowledge support from DST JSPS grant (DST/INT/JSPS/P-238/2017)
and JSPS KAKENHI (grant number JP18K03691).
\def\aj{AJ}%
\def\actaa{Acta Astron.}%
\def\araa{ARA\&A}%
\def\apj{ApJ}%
\def\apjl{ApJ}%
\def\apjs{ApJS}%
\def\ao{Appl.~Opt.}%
\def\apss{Ap\&SS}%
\def\aap{A\&A}%
\def\aapr{A\&A~Rev.}%
\def\aaps{A\&AS}%
\def\azh{AZh}%
\def\baas{BAAS}%
\def\bac{Bull. astr. Inst. Czechosl.}%
\def\caa{Chinese Astron. Astrophys.}%
\def\cjaa{Chinese J. Astron. Astrophys.}%
\def\icarus{Icarus}%
\def\jcap{J. Cosmology Astropart. Phys.}%
\def\jrasc{JRASC}%
\def\mnras{MNRAS}%
\def\memras{MmRAS}%
\def\na{New A}%
\def\nar{New A Rev.}%
\def\pasa{PASA}%
\def\pra{Phys.~Rev.~A}%
\def\prb{Phys.~Rev.~B}%
\def\prc{Phys.~Rev.~C}%
\def\prd{Phys.~Rev.~D}%
\def\pre{Phys.~Rev.~E}%
\def\prl{Phys.~Rev.~Lett.}%
\def\pasp{PASP}%
\def\pasj{PASJ}%
\def\qjras{QJRAS}%
\def\rmxaa{Rev. Mexicana Astron. Astrofis.}%
\def\skytel{S\&T}%
\def\solphys{Sol.~Phys.}%
\def\sovast{Soviet~Ast.}%
\def\ssr{Space~Sci.~Rev.}%
\def\zap{ZAp}%
\def\nat{Nature}%
\def\iaucirc{IAU~Circ.}%
\def\aplett{Astrophys.~Lett.}%
\def\apspr{Astrophys.~Space~Phys.~Res.}%
\def\bain{Bull.~Astron.~Inst.~Netherlands}%
\def\fcp{Fund.~Cosmic~Phys.}%
\def\gca{Geochim.~Cosmochim.~Acta}%
\def\grl{Geophys.~Res.~Lett.}%
\def\jcp{J.~Chem.~Phys.}%
\def\jgr{J.~Geophys.~Res.}%
\def\jqsrt{J.~Quant.~Spec.~Radiat.~Transf.}%
\def\memsai{Mem.~Soc.~Astron.~Italiana}%
\def\nphysa{Nucl.~Phys.~A}%
\def\physrep{Phys.~Rep.}%
\def\physscr{Phys.~Scr}%
\def\planss{Planet.~Space~Sci.}%
\def\procspie{Proc.~SPIE}%
\def\JCSFT{J.~Chem.~Soc.~Faraday~Trans.}%
\let\astap=\aap
\let\apjlett=\apjl
\let\apjsupp=\apjs
\let\applopt=\ao

\bibliography{mridu}{}
\bibliographystyle{plainnat}



\end{document}